\documentclass[a4paper,12pt]{article}

\usepackage{graphicx}
\usepackage{amssymb}
\usepackage{amsmath}
\usepackage[T1]{fontenc}
\usepackage{color}
\usepackage[dvipsnames]{xcolor}
\usepackage[normalem]{ulem}

\usepackage[varg]{txfonts}	% fonts for integrals etc

\def\url#1{\textcolor{blue}{\underline{#1}}}	% web references
\definecolor{oneblue}{rgb}{0,0,0.65}
\definecolor{onered}{rgb}{0.65,0,0}

\def\url#1{\textcolor{blue}{\underline{#1}}}	% web references
\def\email#1{\textcolor{blue}{#1}}	% email address
\def\b#1{\textcolor{oneblue}{#1}} 
 \large\normalsize	 %double
\begin{document}

\title{ Kramers-Kronig relations and the
  properties of conductivity and permittivity in heterogeneous media}

\author{Claude B\'edard et Alain Destexhe\\UNIC, CNRS, Gif sur Yvette,
  France \\ \email{destexhe@unic.cnrs-gif.fr}}

\maketitle

\section*{Abstract\label{abstract}}

The macroscopic electric permittivity of a given medium may depend on
frequency, but this frequency dependence cannot be arbitrary, its real
and imaginary parts are related by the well-known {\it Kramers-Kronig}
relations.  Here, we show that an analogous paradigm applies to the
macroscopic electric conductivity.  If the causality principle is
taken into account, there exists Kramers-Kronig relations for
conductivity, which are mathematically equivalent to the Hilbert
transform.  These relations impose strong constraints that models of
heterogeneous media should satisfy to have a physically plausible
frequency dependence of the conductivity and permittivity.  We
illustrate these relations and constraints by a few examples of known
physical media. These extended relations constitute important
constraints to test the consistency of past and future experimental
measurements of the electric properties of heterogeneous media.

\section{Introduction}

The theory of electromagnetism can be applied to complex media like
inhomogeneous materials or biological tissue.  A first approach to
such media is to explicitly consider their microscopic structure, and
the associated variations of electric conductivity or permittivity,
but this approach requires a detailed mapping of these electric
parameters and include this in detailed simulations.  Another approach
is to use a mean-field electromagnetic theory by considering scales
larger than the typical scales of inhomogeneities in the medium.  In
this case, the mean-field theory relates to macroscopic measurements
of conductivity and permittivity.  For example, in neural tissue,
macroscopic measurements of these quantities were done in a number of
studies~\cite{Gabriel,Logothetis,gomes2016,Miceli} who measured the
frequency dependence of electric parameters in different conditions
(reviewed in ref.~\cite{ BedDes2017}).  Theoretical work showed that
this frequency dependence can be accounted by physical phenomena such
as ionic diffusion or cell polarization~\cite{BedDes2009,BedDes2011a}.
In the present paper, we investigate the relations between the
apparent macroscopic conductivity and permittivity\footnote{Here,
  ``apparent'' refers to the values that can be measured
  experimentally.} and show that they cannot take arbitrary values but
are strongly constrained to be consistent with Maxwell's theory of
electromagnetism.

An important aspect of the equations of electromagnetism when applied
to media with a given electric permittivity \b{$\varepsilon
  (\vec{r},t)$}, is that the real \b{$\varepsilon'$} and imaginary
\b{$\varepsilon''$} parts of the complex Fourier transform of
\b{$\varepsilon (\vec{r},\omega)$} cannot take independent values, but
they are linked together by a set of mathematical relations called
{\it Kramers-Kronig relations} \cite{Landau,Kronig,Forster}.  These
relations are a direct consequence of the principle of causality,
namely that the future cannot influence the past~\cite{Landau}, and
were shown to be equivalent to the Hilbert transform in a perfect
dielectric \cite{Appel,Klitz}.

Here, we will re-examine these Kramers-Kronig relations and their
equivalence to Hilbert transforms in heterogeneous media.  The goal is
to determine the constraints that the frequency dependence of electric
conductivity and permittivity must satisfy.  We will also consider the
application of this formalism to a few concrete examples.

\section{Theory\label{sec1}}

We begin by setting the framework of the present study, by outlining
the electromagnetic theory used here, and how to apply it to
heterogeneous media.  We next consider the Kramers-Kronig relations
within this framework.

\subsection{General framework}

To derive a formalism applicable to heterogeneous media, we
consider linear media within the \textit{electric quasi-static
  approximation} in mean-field{\footnote{Note that there also exists a
    magnetic quasi-static approximation, where electromagnetic
    induction is not negligible, but the displacement current is
    rather neglected.  The latter approximation applies when
    \b{$\parallel \vec{E}
      \parallel~ <<~c \parallel
      \vec{B}\parallel$}}\cite{Bellac,Rousseaux}, which corresponds to
  a physical situation where electromagnetic induction can be
  neglected.  This is the case for neural tissue, where we have an
  excellent approximation of the electric field if we assume \b{
\begin{equation}
\left \{
\begin{array}{cccc}
\nabla\cdot\bf{\vec{D}} &=&   ~~~\bf{\rho^{free}} \\
\nabla\times \bf{ \vec{E}} &=& 0
\end{array}
\right .
\label{eq1}
\end{equation}} 
where the fields \b{$\bf{\vec{E}},~\bf{\vec{D}},~\bf{\rho^{free}}$} are 
mean-fields over a base volume \b{$\mathcal{V}$} \cite{BedDes2011a}. 
By definition, we have \b{$\textbf{X}= < X >{\big |}_{\mathcal{V}}$} }.

In this approximation of the Maxwell-Heaviside equations, the electric
field and magnetic induction are such that \b{$\parallel \bf{\vec{E}}
  \parallel~ >>~c \parallel \bf{\vec{B}}\parallel$} where \b{$c$} is the
velocity of electromagnetic waves \cite{Bellac}.  Thus, in this 
approximation, we can calculate \b{$\vec{\bf{E}}$} independently of
\b{$\vec{\bf{B}}$} and \b{$\vec{\bf{H}}$} if we know the linking
equation between \b{$\vec{\bf{D}}$} and \b{$\vec{\bf{E}}$}.

On the other hand, the link between the electric displacement field
and magnetic field always hold because we have \b{
\begin{equation}
\left \{
\begin{array}{cccc}
\nabla\cdot\vec{\bf{B}} &=&   0\\
\nabla\times \vec{\bf{B}} &=& \mu_o~[\vec{\bf{j}}^{free} + \frac{\partial \vec{\bf{D}}}{\partial t}]
\end{array}
\right .
\label{eq2}
\end{equation}}
where \b{$\mu_o$} is the magnetic permeability of vacuum.  We do not
consider here the situation where this permeability would be different
from vacuum, which normally should be a good approximation because of
the absence of large amounts of ferromagnetic, paramagnetic or
diamagnetic materials in neural tissue.

Moreover, if we restrict to isotropic media (when the inhomogeneity of 
the medium similarly affects all directions), we have:
\b{
\begin{equation}
\left \{
\begin{array}{ccccc}
\vec{\bf{D}}(\vec{r},t) &=& \int_{-\infty}^{+\infty}\bf{\varepsilon}(\vec{r},\tau)~\vec{\bf{E}}(\vec{r},t-t )~d\tau & (a)\\
\\
\vec{\bf{j}}^{free}(\vec{r},t) &=& \int_{-\infty}^{+\infty}
\bf{\sigma}_e(\vec{r},\tau)~\vec{\bf{E}}(\vec{r},t-\tau )~d\tau & (b)
\end{array}
\right .
\label{eq3}
\end{equation}}
where \b{$\bf{\varepsilon}(\vec{r},t)$} and \b{$\bf{\sigma}_e(\vec{r},t) $} are real-valued functions.

This formalism applies to neural tissue, which can be considered as an
heterogeneous isotropic medium within a mean-field context,
when the base volume is sufficiently large (\b{$>1~\mu m^3$}; see
ref.~\cite{BedDes2011a}).  The linking equations imply that the values
of \b{$\vec{\bf{D}}$} and \b{$ \vec{\bf{j}}^{~free}$} at a given time
depend on the past values of the electric field in general, which can
be seen as a kind of ``memory''.  The only way to avoid such a memory
is to assume
\b{$\bf{\sigma}_e(\vec{r},t)=\bf{\sigma}_c(\vec{r})~\delta(t)$} and
\b{$\bf{\varepsilon}(\vec{r},t)=\bf{\varepsilon}_c(\vec{r})~\delta(t
  )$} where \b{$\bf{\sigma}_c$} et \b{$\bf{\varepsilon}_c$} are not
time-dependent.  Note that such a memory is equivalent to having a
frequency dependence of the electric permittivity and/or conductivity,
when we formulate the problem in Fourier frequency space. In this
case, the relations~(\ref{eq3}) imply \b{$\vec{\bf{D}}(\vec{r},\omega
  ) =\bf{\varepsilon}(\vec{r},\omega ) \vec{\bf{E}}(\vec{r},\omega )$}
et \b{$\vec{\bf{j}}(\vec{r},\omega ) =\bf{\sigma}_e(\vec{r},\omega )
  \vec{\bf{E}}(\vec{r},\omega )$}.

Finally, recalling that \b{$ \nabla \cdot \vec{\bf{D}}$} allows us to
calculate the free charge density in a region \b{$\mathcal{D}$}, we
have: \b{
\begin{equation}
\bf{Q}^{free}(t) = \iiint\limits_{\mathcal{D}}\nabla\cdot \vec{\bf{D}}~(\vec{r},t) ~ dxdydz
\label{eq5}
\end{equation}}
The electric permittivity \b{$\varepsilon$} measures the amount of 
free charges in a given region.  The higher the density, the larger the 
permittivity.

\subsection{Electric conductivity and permittivity in a mean-field
  model of isotropic media \label{sec1.1}}
  
In the following, we will re-examine the Kramers-Kronig relations in
the case of an heterogeneous and conductive medium, where the
electric field is time dependent.  In a first step, we show that the
Kramers-Kronig relations are equivalents to the Hilbert transform if
we apply them to \b{$\omega\bf{\varepsilon}(\vec{r},\omega)$} instead
of \b{$\bf{\varepsilon}(\vec{r},\omega)$}.  In a second step, we show
that we have the same relations for the conductivity
\b{$\bf{\sigma}_e$}.

\subsubsection{General expression for the absolute electric
  permittivity \label{sec1.1.1}}

According to relations~(\ref{eq3}), the electric displacement field
\b{$\vec{\bf{D}}$} is linked to the resulting electric field
\b{$\vec{\bf{E}}$} by a convolution integral: \b{
\begin{equation}
\vec{\bf{D}}(\vec{r},t ) = \int_{-\infty}^{+\infty} \bf{\varepsilon}(\vec{r},\tau )~\vec{\bf{E}}(\vec{x},t-\tau )~d\tau
=\int_{-\infty}^{+\infty} \bf{\varepsilon}(\vec{x},t-\tau )~\vec{\bf{E}}(\vec{r},\tau )~d\tau
\label{eq4}
\end{equation}}
However, we must also assume that \b{$\bf{\varepsilon }(\vec{x},t
  )=0$} when \b{$\tau <0$} because the future cannot influence the
past (\textbf{causality principle}).  Thus, we can write \b{
\begin{equation}
\vec{\bf{D}}(\vec{r},t ) = \int_{0^-}^{+\infty} \bf{\varepsilon}(\vec{r},\tau )~\vec{E}(\vec{r},t-\tau )~d\tau
\label{eq6}
\end{equation}}
which is always valid in a plausible physical system.

Moreover, in general, we can write
\b{
\begin{equation}
\bf{\varepsilon}(\vec{r},t)  = \varepsilon_o [\delta(t) + \bf{\chi}(\vec{r},t)]
\label{eq7}
\end{equation}}
which is always valid because \b{$\bf{\chi}$} is an arbitrary function
(or distribution).  The parameter \b{$\varepsilon_o = 8.875\times
  10^{-12}~F/m $} is the vacuum permittivity and \b{$\bf{\chi} $} is
the electric susceptibility expressed in temporal space.  \b{$ \chi$}
measures the amount of free charges induced by applying an electric
field in the medium.  Note that this phenomenon will be necessarily
present in a heterogeneous medium such as neural tissue because we
have charge accumulation in membranes when applied to an electric
field.  Note that the phenomenon of charge induction can change the
law of attenuation with distance, as shown before~\cite{ChapLFP2009}.

Thus, we can write
\b{
\begin{equation}
\vec{\bf{D}}~(\vec{r},t ) =\bf{\varepsilon}_o \int_{0}^{+\infty}[ \delta (t) +\bf{\chi }
(\vec{x},\tau )]~\vec{\bf{E}}(\vec{r},t-\tau )~d\tau
=\varepsilon_o \vec{\bf{E}} +\int_{0}^{+\infty} \varepsilon_o\bf{\chi}(\vec{r},
\tau )~\vec{\bf{E}}(\vec{r},t-\tau )~d\tau~~.
\label{eq8}
\end{equation}}
It follows that, if the Fourier transform relative to
\b{$\omega=2\pi\nu$} exists\footnote{Note that the Fourier transform may
  not necessarily exist for all possible mathematical models.
  However, in the case of heterogeneous media like neural tissue, we
  can safely assume that the Fourier transform 
  of electric field and parameters associated to the medium 
  exist because they have a finite duration. Thus, the time integral 
  does not go to infinity, which means it will necessarily converge 
  because the induced electric charge and susceptibility are bounded.}, 
  then we have \b{
\begin{equation}
\vec{\bf{D}} ~(\vec{r},\omega ) 
=\varepsilon_o ~[1 +\bf{\chi }(\vec{x},\omega )]~\vec{\bf{E}}(\vec{r},\omega )
=\bf{\varepsilon }(\vec{r},\omega )~\vec{\bf{E}}(\vec{r},\omega )
\label{eq9}
\end{equation}}
where
\b{
\begin{equation}
\bf{\varepsilon } (\vec{r},\omega )
=\varepsilon_o ~{\Big [}1 +\int_{~0}^{+\infty} \bf{\chi}(\vec{x},\tau )~e^{-i\omega\tau} d\tau ~
{\Big ]}
\label{eq10}
\end{equation}}

We see that the electric permittivity in Fourier frequency space is in
general a complex function, and its variation with respect to that of 
vacuum is such that 
\b{
\begin{equation}
\bar{\Delta} \bf{\varepsilon}=\bf{\varepsilon}-\varepsilon_o = \int_{~0}^{+\infty}\varepsilon_o \bf{\chi}(\vec{x},\tau )~e^{-i\omega\tau} d\tau ~.
\label{eq11}
\end{equation}}

The latter equation implies that we have \b{$\bar{\Delta}\varepsilon
  =0 $} when \b{$\omega \rightarrow \infty $}. Thus, the permittivity
tends to that of vacuum when the frequency tends to infinity, which is
in good agreement with experimental measurements.  The real part of
\b{$\bar{\Delta} \bf{\varepsilon}(\omega) $} is an even function
relative to the frequency \b{$\omega$} and its imaginary part is an
odd function because \b{$\varepsilon_o \bf{\chi}(t) $} is a real
function.

\subsubsection{Analytic continuation of \b{$\omega\varepsilon $} and the
  Kramers-Kronig or Hilbert transform \label{sec1.1.2}}

We have shown that the imaginary part of
\b{$\bar{\Delta}\bf{\varepsilon }(\vec{r},\omega)$} in Fourier
frequency space must be odd and that the real part must be even if we
want the system to be physically plausible.  Moreover, it was shown
that the real and imaginary parts of permittivity are linked by the
Kramers-Kronig relations.  However, we will develop this transform for
\b{$\omega~\bar{\Delta }\varepsilon(\vec{r},\omega) $} instead of
\b{$\bar{\Delta}\varepsilon(\vec{r},\omega)$} to make the formalism
more uniform between conductivity and permittivity.

 \begin{figure}[bht!] 
\centering
\includegraphics[width=10cm]{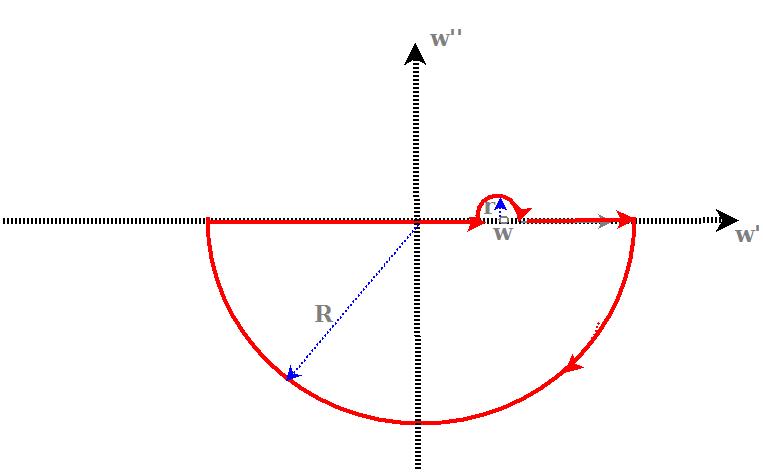}

\caption{Illustration of the integration path in Region
  \b{$\mathbb{P}$}. The singular point \b{$\omega_a=\omega $} is on
  the real segment inside the half circle indicated.  If
  \b{$r\rightarrow 0$} and \b{$R\rightarrow \infty$}, then applying
  the Cauchy integral gives the Hilbert transform. }

\label{fig1}
\end{figure}
 
According to expression (\ref{eq11}), we have
\b{
\begin{equation}
\omega~\bar{\Delta } \bf{\varepsilon} = \int_{~0}^{+\infty}\omega~\varepsilon_o \bf{\chi}(\vec{x},\tau )~e^{-i\omega\tau} d\tau ~
\label{eq12}
\end{equation}}
If we analytically continue the real frequency \b{$\omega $} over the 
complex plane by setting \b{$\omega = \omega' +i \omega''$} then the 
last integral becomes \b{
\begin{equation}
\int_{0}^{+\infty}
\varepsilon_o  \chi(\vec{r},\tau )~\omega e^{-i\omega\tau} d\tau ~
 =\lim\limits_{\delta\rightarrow 0}\int_{\delta}^{1/\delta}
 \varepsilon_o ~\bf{\chi}(\vec{r},\tau ) ~~[~\omega'+i~(\omega''-\delta)~]~ e^{-i\omega'\tau}e^{(\omega''-\delta)\tau}~d\tau
 \label{eq13}
\end{equation}}
where \b{$\delta > 0$}.  The last integral converges when
\b{$\omega''\leq 0$} and tends to zero when \b{$\omega''\rightarrow
  -\infty$}.  It diverges when \b{$\omega''> 0$}.  Note that it is
because we have applied the \textbf{causality principle} that the
integral can converge for \b{$\omega''<0$}; omitting causality would
prevent convergence because one would need to integrate between
\b{$-\infty $} and \b{$+\infty$} (see Eq.~\ref{eq4}).

Thus, the causality principle has the consequence that the function
\b{$\omega \bar{\Delta } \varepsilon $} is holomorphic in the complex
inferior half-plane, as well as on the real axis\footnote{Note that we
  take into account that the permittivity may vary in time, and its
  Fourier transform reflects these time variations.  But one could
  also consider a specific frequency profile of permittivity, and in
  that case, the temporal variations are given by its inverse Fourier
  transform (see ref.~\cite{Landau,Appel}).  In the latter case, one
  must consider the holomorphic transformation in the superior
  half-plane.  }.  If we call this Region \b{$\mathcal{P}$} and apply
the Cauchy integral of a holomorphic function, which gives \b{
 \begin{equation}
\omega~\bar{\Delta } \bf{\varepsilon} (\vec{r},\omega )= \frac{1}{2\pi i}\oint  \frac{\omega_a~\bar{\Delta }\bf{\varepsilon}(\vec{r},\omega_a ) }{\omega_a-\omega} d\omega_a
 \label{eq14}
 \end{equation}}
if the path is closed within Region~\b{$\mathcal{P}$}, and if all
points in this path are all situated at a finite distance from the
origin.  Note that, under this constraint, the function in the
integral has only one singular point at \b{$\omega_a = \omega$}
because \b{$\omega \bar{\Delta} \varepsilon $} is a holomorphic
function inside \b{$\mathcal{P}$}.  This is not the case for \b{$
  \bar{\Delta} \bf{\varepsilon } $} in a conductive medium.

An interesting choice of path is a clockwise along the real axis,
going around the singularity \b{$\omega_a =\omega$} along a
half-circle of radius \b{$\delta$} and a large half-circle of radius
\b{$R=1/\delta$} (see Fig.~\ref{fig1}).  In this case, there is only
the principal value of the integral on the real axis, and we can write
\b{
 \begin{equation}
\omega~\bar{\Delta } \bf{\varepsilon }(\vec{r},\omega ) = -\frac{1}{\pi i}\fint_{-\infty}^{+\infty}  \frac{\omega_a~\bar{\Delta } \bf{\varepsilon }(\vec{r},\omega )}{\omega_a-\omega} d\omega_a
 \label{eq15}
\end{equation}}
For \b{$\omega_a \in \mathbb{R}$} and \b{$|\omega_a|\rightarrow \infty
  $}, we have \b{$|\omega_a~\bar{\Delta } \bf{\varepsilon
  }(\vec{r},\omega )|\rightarrow 0$}. Note that this condition is
equivalent to postulate that \b{$|\bar{\Delta } \bf{\varepsilon
  }(\vec{r},\omega_a )|$} tends to 0 faster than \b{$1/\omega_a$} when
frequency tends to infinity.  This insures that the integral in
Eq.~(\ref{eq15}) converges.  This is obtained because for
\b{$R\rightarrow \infty$} (\b{$\delta\rightarrow 0$}), the integral
over the large circle is zero (see Eq.~\ref{eq10}) while the integral
over the small circle is equal to \b{$-\omega~\bar{\Delta }
  \varepsilon (\vec{r},\omega )/2$} when \b{$\delta \rightarrow 0$}
(see Fig.~\ref{fig1}). We note the principal part of the integral by
\b{$\fint $}

We can separate the real and imaginary parts of 
\b{$\omega~\bar{\Delta } \bf{\varepsilon }(\vec{r},\omega )$}, leading to
\b{
\begin{equation}
\left \{
\begin{array}{ccccc}
\omega~\bar{\Delta } \bf{\varepsilon}'' (\vec{r},\omega )=\mathcal{H}~(\omega~\bar{\Delta } \varepsilon') = \frac{1}{\pi} \fint_{-\infty}^{+\infty} 
\frac{\omega_a~\bar{\Delta } \varepsilon'(\vec{r},\omega_a )}{\omega_a-\omega }
d\omega_a & (a)
\\\\
\omega~\bar{\Delta } \varepsilon' (\vec{r},\omega )=
-\mathcal{H}~(\omega~\bar{\Delta } \varepsilon'') = -\frac{1}{\pi} \fint_{-\infty}^{+\infty} 
\frac{\omega_a~\bar{\Delta } \varepsilon''. (\vec{r},\omega_a )}{\omega_a-\omega }
d\omega_a &(b)
\end{array} 
\right .
\label{eq16}
\end{equation}} 

Note that Eqs.~(\ref{eq16}a) and (\ref{eq16}b) are of opposite sign as
the Kramers-Kronig relations as presented by Landau \&
Lifshitz~\cite{Landau} and Forster \& Schwan~\cite{Forster}. The
reason for this difference is that our analytic continuation of the
electric parameters are holomorphic in the negative half-plane,
instead of the positive half-plane \cite{Appel}\footnote{The starting
  point of Landau \& Lifshitz~\cite{Landau} is to assume that the
  permittivity depends on frequency \textit{a priori}, which defines
  the temporal evolution of the permittivity as the direct Fourier
  transform of the permittivity defined in frequency space.  In our
  approach, we consider that the permittivity is first defined in
  time, while the permittivity in frequency space is given by its
  direct Fourier transform.  These two definitions are mathematically
  equivalent but with a change of sign.  This is why in our formalism,
  we apply the inverse Fourier transform of the permittivity in
  frequency space to obtain the temporal evolution of the
  permittivity.  The two approaches are equivalent, but our approach
  is more appropriate in the case there are time variations of
  electric parameters (such as the opening and closing of ion channels
  ) }.

Consequently, the real and imaginary parts of \b{$\omega\bar{\Delta }
  \varepsilon (\vec{r},\omega )$} are linked by the Hilbert transform.
To calculate the real part from the knowledge of the imaginary part,
one can apply the inverse Hilbert transform, while the direct
transform is used to calculate the imaginary part from the real part.
Note that this transform is totally equivalent to the Kramers-Kronig
relations\footnote{One can recover Kramers-Kronig by considering the
  parity of the real and imaginary parts.}

Finally, we can write
\b{
\begin{equation}
\omega~\bar{\Delta } \bf{\varepsilon}=\omega~\bar{\Delta } \bf{\varepsilon}' + i~\mathcal{H}[\omega\bar{\Delta }\varepsilon']
=-\mathcal{H}[ \omega~\bar{\Delta } \bf{\varepsilon}''] + i~\omega\bar{\Delta }\varepsilon''
\label{eq17}
\end{equation}}
We can see that the phase \b{$\phi$} of \b{$\omega~\bar{\Delta } 
\bf{\varepsilon}$} is such that we have: 
\b{
$tan(\phi)=\frac{\mathcal{H}[\omega\bar{\Delta }\varepsilon']}{\omega~\bar{\Delta } \bf{\varepsilon}'}=
-\frac{\omega\bar{\Delta }\varepsilon''}
{\mathcal{H}[\omega~\bar{\Delta } 
\bf{\varepsilon}'']}
$}. 

Thus, the phase is completely determined when either the real or
imaginary part of \b{$\omega~\bar{\Delta } \bf{\varepsilon}$} is
known.  
\subsection{Electric conductivity within the quasi-static approximation
  \label{sec1.2} }

We now consider the temporal and frequency dependence of the electric
conductivity in linear media within the electric quasi-static
approximation.  In the first section below, we review the constraints
that must be satisfied in Fourier frequency space on the electric
conductivity to simulate a physically plausible system.  We show that
the electric conductivity also obeys to a Hilbert transform, similar
to that shown above for \b{$\omega\bar{\Delta}\varepsilon$}.

\subsubsection{Electric conductivity and free-charge current in linear
  electromagnetism \label{sec1.2.1}}

It is well known that the most general relation between the free-charge
current density \b{$\vec{\bf{j}}^{free}(\vec{r},t)$} and the electric
field \b{$\vec{\bf{E}}(\vec{r},t)$} in linear electromagnetism is given
by the convolution integral:
 \b{
\begin{equation}
\vec{\bf{j}}^{free}(\vec{x},t) = 
\int_{-\infty}^{+\infty} \bf{\sigma}_e(\vec{r},t-\tau )~~\vec{\bf{E}}(\vec{r},\tau )~d\tau=
\int_{-\infty}^{+\infty} \sigma_e(\vec{x},\tau )~~\vec{E}(\vec{x},t-\tau  )~d\tau
\label{eq18}
\end{equation}} 
where \b{$\bf{\sigma }_e \geq 0 $}~
\footnote{This function is necessarily positive or zero, because experiments show that the current density
  is always in the same direction as the electric field.
    The terms under the integral are real functions,
and the function \b{$\bf{\sigma }_e(t)$} is in 
\b{$[\frac{S}{m s}]$}.} 
Expression (\ref{eq1}) can be written as: 
\b{
\begin{equation}
\vec{\bf{j}}^{free}(\vec{r},\omega) = 
\bf{\sigma}_e(\vec{r},\omega)~~\vec{\bf{E}}(\vec{r},\omega)
\label{eq19}
\end{equation}}
in Fourier frequency space. Here, the convolution in temporal space
corresponds to a simple product in frequency space. Note that the
function \b{$\bf{\sigma}_e(\vec{r},\omega)$} depends on the value of the
electric field when the relation between the current density and
eclectic field is nonlinear, but this dependence vanishes if
the system is linear.

Finally, Ohm's law corresponds to the simplest model expressed by
Eqs.~(\ref{eq17}, \ref{eq18}). In this particular case, we have
\b{$\bf{\sigma}_e(\vec{x},t-\tau)=\bf{\sigma}(\vec{x} )~\delta(t-\tau)
  $} where \b{$\bf{\sigma} $} does not depend on time.  \b{$\delta$}
is the Dirac distribution.  This law corresponds to an idealized
physical system without memory (see Section~\ref{sec1}), were the work
produced by the electric field on the free charges dissipates almost
instantaneously\footnote{``Almost instantaneously'' means that the
  dissipation of the energy brought by the electric field is produced
  at \b{$0^+=0 + |dt|$}.} in the system. Note we also have in this
case \b{$\vec{\bf{j}}^{free} (\vec{r},t) = \bf{\sigma}(\vec{r}) ~
  \vec{\bf{E}} (\vec{r},t)$} in temporal space, and we have a similar
relation \b{$\vec{\bf{j}}^{free} (\vec{r},\omega) =
  \bf{\sigma}(\vec{r}) ~ \vec{\bf{E}} (\vec{r},\omega)$} in frequency
space. Thus, we have the same algebraic relation between
\b{$\vec{\bf{j}}^{free}$} and \b{$\vec{\bf{E}} $} in both spaces.

\subsubsection{Constraints imposed by the causality principle
  \label{sec1.2.2}}

To be physically plausible, a system must obey the causality
principle.  This principle determines a constraint on the relation
between the free-charge current density \b{$\vec{\bf{j}}^{free}$} and the
electric field \b{$\vec{\bf{E}}$} (within the linear electromagnetic
theory). According to this principle, the future cannot influence the
past, and thus we can write that, at a given time $t$,
\b{$\vec{\bf{j}}^{free}$} and \b{$\vec{\bf{E}}$} are related as \b{
\begin{equation}
\vec{\bf{j}}^{free}(\vec{r},t) = 
\int_{0}^{+\infty} \bf{\sigma}_e(\vec{r},\tau )~~\vec{\bf{E}}(\vec{r},t-\tau )~d\tau
\label{eq20}
\end{equation}}
because the values of the electric field at times greater than \b{$t$}
cannot influence the current density at time \b{$t$}.  Note that this
is equivalent to assume that \b{$\bf{\sigma}_e(\vec{r},\tau)=0$} when
\b{$\tau < 0$} in Eq.~(\ref{eq19}). This constraint is general and
must be included in all mathematical models of conductivity to be
physically plausible.

\subsubsection{Electrical conductivity in a heterogeneous medium
  \label{sec1.2.3} }

We now consider the case of a heterogeneous medium composed of
different cells and various processes immersed in a conductive
fluid, such that the distance between different elements
\b{$\delta_c$} is always greater than zero.  We also assume that the
electric conductivity of this medium tends asymptotically to
\b{$\bf{\sigma}_{\infty }(\vec{r},t)$} when the frequency \b{$\nu$}
tends to infinity.  For high frequencies, there is a portion of free
charges which does not meet any process, because their mean
displacement becomes smaller than \b{$\delta_c$}.  However, for
sufficiently low frequencies, the presence of cells will impact all
free charges, and the conductivity will be affected and will be
different as that of high frequencies.  Thus, the conductivity of a
heterogeneous medium will necessarily be frequency dependent within
some frequency range.

This intuitive explanation can be formulated more quantitatively.
The electric conductivity can be expressed as \b{
\begin{equation}
 \bf{\sigma}_e(\vec{r},t )=  \bf{\sigma}_{\infty}(\vec{r},t)[1+ \bar{\Delta}_{e}(\vec{r},t)]
 \label{eq21}
\end{equation}}
where \b{$\bar{\Delta}_{e}(\vec{r},t) \neq 0$}.  It follows that 
Expression~(\ref{eq4}) can be written as
\b{
\begin{equation}
\vec{j}^{free}(\vec{r},t) =
~~\int_{0}^{+\infty} \sigma_{\infty}(\vec{x},\tau)~~[~1+ \bar{\Delta}_{e}(\vec{x},\tau)~~]~~\vec{E}(\vec{x},t-\tau )~d\tau~
\label{eq22}
\end{equation}}
Taking the Fourier transform, we obtain
\b{
\begin{equation}
\vec{\bf{j}}^{free}(\vec{r},\omega) ={\Big [}~~\bf{\sigma}_{\infty}(\vec{r}, \omega) 
+\int_{0}^{+\infty} \bf{\sigma}_{\infty}(\vec{r}, \tau)~~ \bar{\Delta}_{e}(\vec{r},\tau )  ~e^{-i\omega\tau}~d\tau~ {\Big ]}~~\vec{\bf{E}}(\vec{r},\omega )
\label{eq23}
\end{equation}}
Thus, we can write
\b{
\begin{equation}
\bf{\sigma}_e(\vec{r},\omega )=\bf{\sigma}_{\infty}(\vec{r}, \omega)+\int_{0}^{+\infty}\bf{\sigma}_{\infty}(\vec{r}, \tau)~~
\bar{\Delta}_{e}(\vec{r},\tau ) ~~ e^{-i\omega\tau}~d\tau 
\label{eq24}
\end{equation}}
where the function \b{$\bf{\sigma}_{\infty}(\vec{r},
  \tau)\bar{\Delta}_{e}(\vec{r}, \tau)$} is such that the integral in
Expression~(\ref{eq24}) tends to zero when
\b{$\omega\rightarrow\infty$}. Thus,
\b{$\bf{\sigma}_e(\vec{r},\omega\rightarrow \infty )$} tends to
\b{$\bf{\sigma}_{\infty}(\vec{r}, \omega)$}.  Note that the integral in
the righthand side converges because it is the Fourier transform of
\b{$\bar{\Delta }\bf{\sigma}_e=\bf{\sigma}_e -\bf{\sigma}_{\infty}$}.

\subsubsection{Similar relations between
  \b{$\omega\bar{\Delta}\bf{\varepsilon}$} and \b{$\bar{\Delta}\bf{\sigma}_e$}
  \label{sec1.2.4}}

We know that the imaginary part of \b{$\bar{\Delta}\bf{\sigma}_e$} in
frequency space is an odd function, while the real part is even, if
the model is physically plausible because the time-dependent
conductivity is a real function.  However, these parts could still be
independent of each-other.  We now show that if we apply the causality
principle, the imaginary part of \b{$\bar{\Delta}\bf{\sigma}_e$} (in
frequency space) is completely determined by its real part, via a
Hilbert transform, exactly like \b{$\omega\bar{\Delta}\bf{\varepsilon}
  $}.

If we analytically continue the frequency \b{$\omega $} over 
the complex plane by taking \b{$\omega = \omega' +i \omega''$}, then
the integral in Expression~(\ref{eq23}) becomes \b{
\begin{equation}
\int_{0}^{+\infty}
\bf{\sigma}_{\infty}(\vec{r}, \tau) \bar{\Delta}_{e}(\vec{r},\tau ) ~~ e^{-i\omega\tau}~d\tau
 =\lim\limits_{\delta\rightarrow 0}\int_{\delta}^{1/\delta}
\sigma_{\infty}(\vec{r}, \tau) \bar{\Delta}_{e}(\vec{r},\tau ) ~~ e^{-i\omega'\tau}e^{(\omega''-\delta )\tau}~d\tau
 \label{eq25}
\end{equation}} 
where \b{$\delta > 0$}.  This integral converges when \b{$\omega''<
  0$} and tends to 0 for \b{$\omega''\rightarrow -\infty$}. On the
other hand, it diverges when \b{$\omega''> 0$}. Note that the
principle of causality allows the convergence of this integral when
the imaginary part of \b{$\omega$} is smaller than zero.  This would
not be the case if the causality principle is not used, because one
would need to integrate between \b{$-\infty $} and \b{$+\infty$}

Thus, the situation is completely analogous to the analytic continuation
of \b{$\omega\bar{\Delta }\varepsilon$}, and we can write:
\b{
 \begin{equation}
 \left \{
 \begin{array}{ccccc}
 \bar{\Delta}\bf{ \sigma}_e''
 =\mathcal{H}~(~\bar{\Delta } \bf{\sigma}_e') = \frac{1}{\pi}\fint_{-\infty}^{+\infty} 
  \frac{\bar{\Delta}\bf{\sigma}_e'(\vec{r},\omega_a)}{\omega_a-\omega} d\omega_a~
  & (a)
  \\\\
  \bar{\Delta}\bf{ \sigma}_e' =
  -\mathcal{H}~(~\bar{\Delta } \bf{\sigma}_e'')
  =
  - \frac{1}{\pi}\fint_{-\infty}^{+\infty}   
  \frac{\bar{\Delta}\bf{\sigma}_e''(\vec{r},\omega_a)}{\omega_a-\omega} d\omega_a~
  &(b)
  \end{array}
  \right .
  \label{eq26}
\end{equation}} 
\b{$\bf{\sigma}_e(\vec{r},\omega)=\bf{\sigma}_e'(\vec{r},\omega)+i\bf{\sigma}_e''(\vec{r},\omega)$} and
\b{$\bf{\sigma}_{\infty}=\bf{\sigma}_{\infty}'+i\bf{\sigma}_{\infty}''$}.
\b{$\bf{\sigma}_e'(\vec{r},\omega)$} and \b{$\bf{\sigma}_{\infty}'$} are
respectively the real part of \b{$\bf{\sigma}_e$} and
\b{$\bf{\sigma}_{\infty}$}, while \b{$\bf{\sigma}_e''(\vec{r},\omega)$} and
\b{$\bf{\sigma}_{\infty}''$} are respectively their imaginary part.
Consequently, if we know the conductivity at very high frequencies and
the real part of its frequency dependence, then we can determine the
corresponding imaginary part using the Hilbert transform, and
\textit{vice et versa}.  Finally, note that we apply here the Hilbert
transform to the variation of conductivity relative to
\b{$\sigma_e(\vec{r},\infty)$} to make sure that the integral
converges when the frequency tends to infinity. 

%This problem does not
%occur with the permittivity because it tends to a constant (vacuum
%permittivity) when frequency tends to infinity.

\subsection{Apparent conductivity and permittivity between two
  isopotential surfaces \label{sec1.3}}

Within the electric quasi-static approximation, the electric field can
be expressed as the gradient of the potential, \b{$\vec{\bf{E}} =
  -\nabla \bf{V}$}.  We also know that the free-charge current is not
conserved in general in a heterogeneous medium, for example because
charge accumulation can occur\footnote{In the Debye layers of ions
  around a membrane, for example.}  The density of the generalized
current is such that \b{$\nabla\cdot \vec{\bf{j}}^g =0 $}, where
\b{$\vec{\bf{j}}^g = \vec{\bf{j}}^{~free} +
  \frac{\partial\vec{\bf{D}}}{\partial t}$}.  Note that this does not
represent a stationary law, because \b{$\vec{\bf{j}}^g$} is time
dependent, but it is rather a conservation law \cite{BedDes2013}.  The
generalized current entering a given domain is always equal to the
generalized current exiting that domain, even if charge accumulation
occurs\footnote{Note that this is only true at chemical steady-state,
  when there is no current created by chemical reactions (see
  discussion in ref.~\cite{BedDes2014}).}.

Thus, according to these laws, the generalized current density between
two close-by equipotential surfaces is given by:
\b{
\begin{equation}
\vec{\bf{j}}^g (\vec{r},\omega)= -\gamma (\vec{r},\omega) \nabla \bf{V} (\vec{r},\omega)
=-(\bf{\sigma}_e +i\omega\bf{\varepsilon}) \nabla \bf{V} (\vec{r},\omega)
\label{eq27}
\end{equation}}
where \b{$\bf{\gamma}$} is the admittance of the medium.  \b{$\bf{\sigma}_e$} is
the link between \b{$\vec{\bf{j}}^{free}$} and \b{$\vec{\bf{E}}$} in frequency
space, while\b{$\bf{\varepsilon}$} is the link between \b{$\vec{\bf{E}}$} and
\b{$\vec{\bf{D}}$}.  The latter link allows one to calculate the induced
charge in a given region of the medium by a given electric field,
expressed in Fourier frequency space.  Note that these links are all
complex numbers in general because there can be a non-zero phase,
similar to a capacitance.

 \begin{figure}[bht!] 
\centering
\includegraphics[width=10cm]{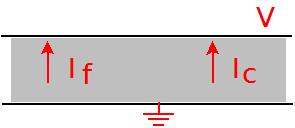}

\caption{Capacitance with a leak current \b{$I_f$}. In Fourier
  frequency space, the voltage difference \b{$V$} and the current
  density \b{$j$} obey \b{$\vec{j}=[\sigma_e +i \omega \varepsilon]
    ~\frac{V}{d} $} when the electric parameters do not depend on
  frequency.  \b{$d$} is the distance between the arms of the
  capacitor. The density of the leak current is given by \b{$j_f =
    \sigma_e V$} that of the capacitive current is \b{$j_c =
    i\omega\varepsilon V$} . We also have \b{$I_f =j_fA$ } and \b{$I_c
    =j_cA$ } where $A$ is the area of the arms.}

\label{fig2}
\end{figure}

If we now assume the following equalities:
 \b{
\begin{equation}
\bf{\gamma}(\vec{r},\infty) =\bf{\sigma}(\vec{r},\infty ) +i\omega\bf{\varepsilon} 
(\vec{r},\infty )  ~~et~~
\bar{\Delta }\bf{\gamma}  =\bf{\gamma}(\vec{r},\omega )-\bf{\gamma}_{\infty}
\label{eq28}
\end{equation}}
Applying Equations~(\ref{eq16}) and (\ref{eq25}), we can write:
\b{
\begin{equation}
\bar{\Delta }\bf{\gamma}=\bar{\Delta }\bf{\gamma}'+i\bar{\Delta }\gamma''  = 
[\bar{\Delta } \bf{\sigma}_e'- \mathcal{H}(\omega\bar{\Delta }\bf{\varepsilon}')]  +i~[\omega\bar{\Delta }\bf{\varepsilon}'+\mathcal{H}(\bar{\Delta }\bf{\sigma}_e')] 
\label{eq29}
\end{equation}}
It follows that
\b{
\begin{equation}
\left \{
\begin{array}{ccccc}
\bar{\Delta }\bf{\gamma}'' = \mathcal{H}~(\bar{\Delta } \bf{\gamma}')
& (a)
\\\\
\bar{\Delta }\bf{\gamma}' = -\mathcal{H}~(\bar{\Delta } \bf{\gamma}'')
& (b)
\end{array}
\right .
\label{eq30}
\end{equation}}
because the Hilbert transform obeys \b{$-\mathcal{H}^2~(F) =F$}. 
We can also write
\b{
\begin{equation}
\bar {\Delta  } \bf{\gamma} = \bf{\bar {\Delta  }\gamma}' + i~ \mathcal{H}~(\bf{\bar {\Delta  }\gamma'})~~
\label{eq31}
\end{equation}}
such that the phase \b{$\phi$} obeys
 \b{\begin{equation}
tan(\phi) = \frac{ \mathcal{H}(\bf{\bar{\Delta}\gamma'})}{\bf{\bar{\Delta}\gamma}'} 
 \label{eq32}
 \end{equation}}

For a sufficiently large base volume, we can always assume that the
conductivity and permittivity do not depend on position.  In this
particular case, we can write \b{
\begin{equation}
\left \{
\begin{array}{cccccc}
I^g &=& \iint\limits_{\mathcal{S}_{eq}}\vec{j}\cdot\hat{n}~dS &
=&<\parallel \vec{j} \parallel >_{\mathcal{S}}S & (a)
\\\\
\Delta V &=& \int_a^b \nabla V \cdot \hat{n}~ds &
= &
\frac{<\parallel \vec{j} \parallel >_{l}}{\gamma(\omega )}l &
 (b)
\end{array}
\right .
\label{eq33}
\end{equation}}
where the current \b{$I^g$} is conserved.  \b{$<\parallel \vec{j}
  \parallel >_{\mathcal{S}}$} is the mean current density over an
equipotential surface and \b{$<\parallel \vec{j}
  \parallel >_{l}$} is the mean current density along a current line
between two equipotential surfaces.  These two means are different in
general, but if we take the curve that goes through the mean current
density \b{$<\parallel \vec{j}
  \parallel >_{\mathcal{S}}$} over each equipotential surface, then
we can say that this curve is a current line.  Consequently, there
exists an equipotential surface and a current line such that
\b{$<\parallel \vec{j} \parallel >_{\mathcal{S}}=<\parallel \vec{j}
  \parallel >_{l}$}.  It follows that, if \b{$A_\mathcal{C}$} is the
area of this surface, and \b{$l_\mathcal{C}$} is the distance between
two isopotential surfaces, then we can write \b{
\begin{equation}
\frac{I^g}{\Delta V}= \gamma (\omega )
~\frac{A_\mathcal{C}}{l_\mathcal{C}}
\label{eq34}
\end{equation}}
This form is identical to that of a plane capacitor which possesses a
leak current, because the real part is non-zero, or a conductance with
a non-negligible capacitive effect, because the imaginary part is
also non-zero.  

If we define the apparent conductivity as \b{$\sigma_{\mathcal{A}}$} 
as the real part of \b{$\bf{\gamma }$}
\b{
\begin{equation}
\sigma_{\mathcal{A}}=\gamma'= \sigma_e'- \omega\varepsilon'' 
\label{eq35}
\end{equation}} 
and the apparent permittivity \b{$\varepsilon_{\mathcal{A}}$} times
the angular frequency \b{$\omega $} as the imaginary part of 
\b{$\bf{\gamma}$}
\b{
\begin{equation}
\omega\varepsilon_{\mathcal{A}}=\gamma''=\omega
\varepsilon'+\sigma_e''~,
\label{eq36}
\end{equation}}
we can then write
where these apparent parameters are real and linked by a Hilbert
transform .  We have
\b{$\omega\bar{\Delta}\varepsilon_{\mathcal{A}} = \mathcal{H}(\bar{\Delta}\sigma_{\mathcal{A}})$}
and
\b{$\bar{\Delta}\sigma_{\mathcal{A}} = -\mathcal{H}(\omega\bar{\Delta}\varepsilon_{\mathcal{A}})$}. 

Thus, the knowledge of one parameter is sufficient to deduce the
other.  For a given frequency, it is always possible to simulate the
current-voltage relation between two isopotential surfaces as a plane
capacitor (see Fig.~\ref{fig2}).  The leak current of this capacitor
is determined by the Hilbert transform of its admittance, or by a
conductance possessing a capacitive effect determined by Hilbert
transform of the admittance. Finally, \b{$\sigma_{\mathcal{A}}$} and
\b{$\varepsilon_{\mathcal{A}}$} are an even function (see
Eqs.~\ref{eq35}) and \ref{eq36}) relative to \b{$\omega $}.

It is important to note that, for the apparent parameters, as defined
in Eqs.~\ref{eq35} and \ref{eq36}, we have the following properties.
First, the apparent conductivity is equal to the electric conductivity
when \b{$\omega =0$}, or when the imaginary part of the electric
permittivity is zero.  Second, the apparent permittivity becomes
infinitely large for \b{$\omega=0$} when the imaginary part of the
electric conductivity is different from zero.  Third, if the imaginary parts of both
electric conductivity and permittivity are zero, then the apparent
parameters are identical to electric parameters.  Finally, if the
physical effects associated to the electric permittivity (density of
induced charges) are negligible compared to that of electric
conductivity, then the apparent conductivity is approximately given by
the real part of electric conductivity, and the apparent permittivity
is approximately to the imaginary part of electric conductivity
divided by \b{$\omega$}.

\section{Applications\label{sec2}}

In this section, we consider models of different known media to
illustrate the consequences of the relations outlined in the
theoretical part.  All models considered are built within the
quasistatic approximation of the linear electromagnetic theory of
Maxwell-Heaviside.  In other words, we do not consider physical
phenomena associated to electromagnetic induction
(\b{$\nabla\times\vec{\bf{E}}=0$}).  In such conditions, we can apply
the Kramers-Kronig relations (or Hilbert transform) over the
approximated electric parameters~\footnote{By ``approximated''
  parameters, we mean the electric parameters within the quasistatic
  electric approximation. They constitute an excellent approximation
  for low frequencies, \b{$< \sim 10KHz$}.  }.

\subsection{Constraints on the conductivity given by the
  Kramers-Kronig relations}
\label{sec2.1}

\subsubsection{First example \label{Ex1} }

As a first example, we consider an isotropic medium where the base
volume of the mean-field does not depend on position.  We suppose that
the medium is such that \b{$\sigma_e$} depends on frequency with a
null imaginary part, and that
\b{$\lim\limits_{\omega\rightarrow\infty} \sigma_e=k$} is real and
frequency independent.  This model would correspond to a heterogeneous
and isotropic medium where the conductivity depends on frequency
\b{$\nu$} according to a law \b{$\sigma_e(\omega=2\pi\nu
  )=\sigma_e'(\omega ) + i\sigma_e''(\omega )=\sigma_e'(\omega )$}.

In this case, according to Eq.~(\ref{eq26}b), we must have
\b{
\begin{equation}
 \sigma_e'(\omega)=\sigma(\infty ) =k
 \label{eq38}
\end{equation}}
because \b{$\bar{\Delta} \sigma_e' =0 $} when \b{$\bar{\Delta}
  \sigma_e'' =0 $}.  Therefore, according to Kramers-Kronig relations,
\b{$\sigma_e'$} would not depend on frequency, which is contradictory
with the initial hypothesis.

Thus, a model of this kind is not acceptable, because we have a
contradiction with the causality principle because the Kramers-Kronig
relations are a direct consequence of this principle in a linear
system.

For example, it is impossible to have a conductivity law of the form
\b{$\sigma_e = \omega^xe^{-\omega/c} +k$} where \b{x} is arbitrary.
One must add an imaginary part, which implies that the phase is
non-zero.  Thus, a model with a frequency-dependent conductivity but a
null phase is physically impossible in a model where conductivity
becomes resistive or if the phase becomes zero at high frequencies .
Note that, in general, the electric conductivity is not necessarily
real for large frequencies.  It was shown before that the linear
approximation of a physical system with ionic diffusion gives an
electric conductance of the form
\b{$\sqrt{\omega}(a+ib)$}~\cite{BedDes2011a}.  Thus, the asymptotic
behavior of the electric conductivity at high frequencies is very
different from the electric permittivity, because the latter tends to
the permittivity of vacuum, which is real.

For example, let us consider the model examined by Miceli et
al.~\cite{Miceli} of a medium in which the electric conductivity is
frequency-dependent, but the permittivity is constant. According to
above, such a medium is physically impossible, because it would
violate the principle of causality.  It is important to note that the
standard model of a resistive extracellular medium, the capacitive
effects are neglected.

The Miceli model is asymptotically resistive at high frequencies
(\b{$\lim\limits_{\omega\rightarrow\infty} \sigma_e=k$}), and thus, we
can conclude that, if the conductivity depends on frequency, one must
necessarily have both real and imaginary parts different from zero for
the model to be physically plausible.

\subsubsection{Second example \label{Ex2} }

In this section, we suppose that the apparent macroscopic permittivity 
 \b{$\varepsilon_{\mathcal{A}}$} between two equipotential surfaces
is given by
\b{
\begin{equation}
\begin{array}{cccccc}
\varepsilon_{\mathcal{A}} &=& \varepsilon_o + \frac{\kappa }{| \omega|^{2+x} } 
\end{array}
\label{eq39}
\end{equation}}
where \b{$ x,\kappa \in \mathbb{R} $} and \b{$ x\geq -1 $}, and \b{$
  \kappa\geq 0 $} and frequency-independent. Note that this model
respects the parity of \b{$\varepsilon_{\mathcal{A}}$} (see
Eq.~\ref{eq36}) because we have \b{$\varepsilon_{\mathcal{A}}(
  +\omega)= \varepsilon_{\mathcal{A}}( -\omega)$}.  This simple model
corresponds to a physical situation where the apparent permittivity
diminishes for increasing frequency, and approaches that of vacuum for
very high frequencies.  This situation is often encountered in
heterogeneous media such as biological tissues (see for example, the
experimental measurements of Gabriel et al.~\cite{Gabriel}).

Note that if the apparent permittivity \b{$\varepsilon_{\mathcal{A}}$}
tends to that of vacuum, then the electric conductivity \b{$\sigma_e$}
must necessarily have a zero phase when frequency tends to infinity
(see Eq.~\ref{eq36}) because the real part of the electric permittivity
\b{$\varepsilon$} tends to that of vacuum.

By applying Eq.~(\ref{eq30}b) and assuming that \b{$ y =
  \frac{\omega_a }{ w } $}, we obtain:
\b{
\begin{equation}
\bar{\Delta }\sigma_{\mathcal{A}}(\omega) = -\mathcal{H}~( \bar{\Delta}\omega\varepsilon_{\mathcal{A}})
= -\frac{2~\kappa}{\pi\omega^{1+x}} \fint_{~0}^{+\infty}  \frac{1}{(y^2-1)y^{x}}dy   =  -\frac{\kappa_1}{\omega^{1+x}}
\label{eq40}
\end{equation}}
where we have
\b{
\begin{equation}
\kappa_1 = \frac{2~\kappa}{\pi} \fint_{~0}^{+\infty}  \frac{1}{(y^2-1)y^{x}}dy
\label{eq41}
\end{equation}} 
Consequently, the phase of the admittance variation is such that
\b{
\begin{equation}
tan {(\phi ) } = \frac{\bar{\Delta}(\omega\varepsilon_{\mathcal{A}})}{\bar{\Delta }(\sigma_{\mathcal{A}})}= -\frac{\kappa }{\kappa_1} 
\label{eq42}
\end{equation}}
when \b{$\omega > 0$} and \b{$+\frac{\kappa }{\kappa_1}$} for
\b{$\omega <0$}. Note that \b{$\kappa_1 $} depends on the value of the
exponent \b{$x$}.  Also note that the phase of the admittance
variation \b{$\Delta \gamma $} does not depend on frequency in this
type of model.

We now calculate the phase for different values of the exponent \b{$x$}. 
The principal part of the integral in expression (\ref{eq41}) is explicitly
given by the following expression:
\b{
\begin{equation}
\kappa_1 =  \lim\limits_{\delta \rightarrow 0}~{\Big [ } \int_{\delta}^{1-\delta}
\frac{1}{(y^2-1)y^{x}}dy
+\int_{1+\delta}^{1/\delta}\frac{1}{(y^2-1)y^{x}}dy~ {\Big ]}
\label{eq43}
\end{equation}}
where \b{$\delta >0$}.

\paragraph{ Case x=0.}
For \b{$x=0$}, Eq.~(\ref{eq43}) becomes
\b{
\begin{equation}
\kappa_1 =  \frac{2~\kappa}{\pi}~\lim\limits_{\delta \rightarrow 0}~
In~ {\Big [}\frac{-\delta}{2-\delta}\cdot\frac{\delta+1}{\delta-1}\cdot
\frac{1-\delta}{1+\delta}\cdot\frac{2+\delta}{\delta}
{\Big ]}=0
\label{eq44}
\end{equation}}

It follows that the phase of the admittance variation is given by
\b{$\phi = -\frac{\pi}{2}$} for (\b{$\omega >0$}, and \b{$\phi =
  \frac{\pi}{2}$} for (\b{$\omega <0$}.  

Moreover, we see that the variations of apparent conductivity with
frequency are such that \b{$\bar{\Delta} \sigma_{\mathcal{A}}=0$}.
Thus, in this model, the apparent conductivity is equal to the real
part of the electric conductivity because, for infinite frequencies,
the imaginary part of electric permittivity \b{$\varepsilon"$} tends
to zero and \b{$\varepsilon'$} tends to that of vacuum.  Thus, we
have \b{$\varepsilon_{\mathcal{A}}=\varepsilon_o
  +\frac{\kappa}{|\omega|^2} $} and
\b{$\sigma_{\mathcal{A}}=\sigma_e'(\omega \rightarrow\infty) $}.  Consequently,
if we assume that the medium is resistive for very high frequencies,
then we see that the phase is necessarily frequency-dependent because
the phase of the admittance \b{$\gamma$} is given by \b{$tan (\phi) =
  \frac{\omega\varepsilon_0 +\frac{\kappa}{\omega }}{\sigma_e'}$} for
\b{$\omega>0 $}.  

\paragraph{ Case \b{$x= \pm 1/2$} .}
In this case, the primitives of the integrals in expression~(\ref{eq43})
are:
\b{
\begin{equation}
\frac{1}{2} In~ {\Big [}\frac{1-\sqrt{y}}{1+\sqrt{y}}{\Big ]} \mp tg^{-1}\sqrt{y}
\label{eq45}
\end{equation}}
It follows that
\b{
\begin{equation}
\kappa_1 =  \frac{2~\kappa}{\pi}~\lim\limits_{\delta \rightarrow 0}~
In~ {\Big [}\frac{1-\sqrt{1-\delta}}{1+\sqrt{1+\delta}}\cdot\frac{1+\sqrt{\delta}}{1-\sqrt{\delta}}\cdot
\frac{1-\sqrt{1/\delta}}{1+\sqrt{1/\delta}}\cdot\frac{1+\sqrt{1+\delta}}{1-\sqrt{1+\delta}}{\Big ]}+  t =  t
\label{eq46}
\end{equation}}
where \b{$$
t=\mp \frac{2~\kappa}{\pi}\lim\limits_{\delta \rightarrow 0}~[tg^{-1}\sqrt{1-\delta}-tg^{-1}\sqrt{\delta}+tg^{-1}\sqrt{1/\delta}-tg^{-1}\sqrt{1+\delta}]
=\mp \kappa
$$}

Thus, the phase (Eq.~\ref{eq42}) of the admittance variation is given
by \b{$\phi =\pm \frac{\pi}{4}$}, because we have
\b{
\begin{equation}
tan {(\phi ) } = \frac{\bar{\Delta}(\omega\varepsilon_{\mathcal{A}})}{\bar{\Delta }(\sigma_{\mathcal{A}})}= \pm 1 
\label{eq47}
\end{equation}}
Consequently, the admittance variation \b{$\bar{\Delta }\gamma_{\mathcal{A}} 
$} is such that
\b{$
\bar{\Delta }\gamma_{\mathcal{A}}=\bar{\Delta }(\sigma_{\mathcal{A}})(1 - i)
$} and \b{$
\bar{\Delta }\gamma_{\mathcal{A}}=\bar{\Delta }(\sigma_{\mathcal{A}})(1 + i)
$}
when we have, respectively,
\b{
$ 
\varepsilon_{\mathcal{A}} = \varepsilon_o + \frac{\kappa }{| \omega|^{3/2} }
$}
and
\b{
$ 
\varepsilon_{\mathcal{A}} = \varepsilon_o + \frac{\kappa }{| \omega|^{5/2} }
$}

This result is interesting for the following reason: we note that if
we consider the case where the exponent equals \b{$-3/2$} for
\b{$\bar{\Delta} \varepsilon_{\mathcal{A}} $}, then we have
\b{$\bar{\Delta} \sigma_{\mathcal{A}} < 0$}.  This corresponds to a
situation where conductivity increases with frequency.  On the other
hand, if the exponent equals \b{$-5/2$}, then the apparent
conductivity decreases with frequency.  The first case seems in
agreement with experimental measurements (see \cite{Gabriel}).

It follows that, for the first case (exponent of -3/2), we can write:
\b{
\begin{equation}
\gamma_{\mathcal{A}}=\bar{\Delta }\gamma_{\mathcal{A}}
 +\gamma_{\mathcal{A}}( \omega=\infty )= 
 \bar{\Delta }(\sigma_{\mathcal{A}})(1 - i)  +\gamma_{\mathcal{A}}( \omega=\infty )
 \label{eq48}
\end{equation}}

We see that if the apparent admittance tends asymptotically to a
constant \b{$C$} (independent of frequency) for high frequencies, then
we have \b{$C \approx \bar{\Delta }(\sigma_{\mathcal{A}})$}, so there
will be a non-negligible phase at low frequencies.  We recall that, by
definition, we have \b{$\lim\limits_{\omega\rightarrow \infty}
  \bar{\Delta }(\sigma_{\mathcal{A}} )=0$ } (see Eq.~\ref{eq28}).

We also see that if the product \b{$\omega \varepsilon$} is negligible
compared to electric conductivity, and if the latter tends to a
Warburg impedance for high frequencies (as in ref.~\cite{Pham}), then,
according to the definition of apparent parameters (Eqs.~\ref{eq35}
and \ref{eq36}), we can write: \b{
\begin{equation}
\gamma_{\mathcal{A}}=\gamma_e =(\bar{\Delta }(\sigma_{\mathcal{A}})+ k)\sqrt{\omega}~(1\pm i)
\label{eq49}
\end{equation}}
where \b{$\gamma_e$} is the electric admittance.

\paragraph{Case x arbitrary and greater than -1.}

For an arbitrary value of \b{$x$}, we can write
\b{
\begin{equation}
tan {(\phi ) } = \frac{\bar{\Delta}(\omega\varepsilon_{\mathcal{A}})}{\bar{\Delta }(\sigma_{\mathcal{A}})} = cst~
\label{eq50}
\end{equation}}
such that we can write
\b{
\begin{equation}
\gamma_{\mathcal{A}}=\bar{\Delta }\gamma_{\mathcal{A}}
 +\gamma_{\mathcal{A}}( \omega=\infty )= 
 \bar{\Delta }(\sigma_{\mathcal{A}})(1 +tan {(\phi ) }~ i)  +\gamma_{\mathcal{A}}( \omega=\infty )
 \label{eq51}
 \end{equation}}
where the phase \b{$\gamma_{\mathcal{A}}$} is frequency dependent
because \b{$ \bar{\Delta }(\sigma_{\mathcal{A}})$} depends on
frequency.

Finally, we see that if the electric permittivity varies according
to the following law: 
\b{
\begin{equation}
\begin{array}{cccccc}
\varepsilon_{\mathcal{A}} &=& \varepsilon_o +\sum\limits_{j=1}^N \frac{\kappa_j }{| \omega|^{2+x_j} } 
\end{array}
 \label{eq52}
\end{equation}}
where \b{$x_j>-1$}, then we can calculate the phase of the apparent
electric admittance, by fitting the expression (\ref{eq52}) to the
measured electric permittivity.  Note that in these examples, this
phase is non-negligible when the electric or apparent parameters
(admittance, conductivity, permittivity) are frequency dependent.

\section{Discussion\label{sec3}}

In this paper, we have re-examined electromagnetism theory in
heterogeneous media, in particular focusing on the Kramers-Kronig
relations, following the Landau \& Lifshitz formalism~\cite{Landau}.
Our main finding is that, similar to the well-known Kramers-Kronig
relations linking the real and imaginary parts of the electric
permittivity~\cite{Landau}, one can also derive similar relations for
electric conductivity.  Thus, in heterogeneous media, the two electric
parameters obey symmetric dependencies in Fourier frequency space.
This finding is general, and also applies to a homogeneous medium
(which is a particular case of heterogeneous media).  We discuss below
the significance of these results.

As a first example, we considered the model of a medium where electric
conductivity was assumed to be frequency-dependent, but with a
constant permittivity~\cite{Miceli}.  We showed that, according to
Kramers-Kronig relations, this model is physically impossible, as it
would violate the principle of causality.  This models is also
asymptotically resistive at high frequencies
(\b{$\lim\limits_{\omega\rightarrow\infty} \sigma_e=k$}), and thus, we
can conclude that, if the conductivity depends on frequency, one must
necessarily have both real and imaginary parts different from zero for
such a model to be physically plausible.  It is important to note that
the NEURON simulator~\cite{Hines} used for such simulations cannot
deal with complex electric parameters (non-negligible phase), so
cannot be used to simulate physically-plausible non-resistive
situations.

In a second example, we considered experimental measurements
suggesting that the extracellular medium around neurons is
non-resistive~\cite{gomes2016,BedDes2017}.  We showed here that these
measurements are consistent with Kramers-Kronig relations, and the
principle of causality. However, other measurements suggesting
resistive media~\cite{Logothetis,Miceli}, are also consistent with
Kramers-Kronig relations.  All these experimental measurements are
thus physically plausible and self-consistent.  On the modeling point
of view, the models of non-resistive
media~\cite{gomes2016,BedDes2009,BedDes2011a,BedDes2013}, as well as
those of resistive media~\cite{Miceli}, are all consistent with
Kramers-Kronig relations as well.  One notable exception is the
non-resisistive models examined in \cite{Miceli}, which are
non-plausible, as discussed above.  Note that ionic diffusion was
proposed as a mechanism to explain the non-resistive
measurements~\cite{gomes2016,BedDes2017,BedDes2009}, and according to
this mechanism, the phase of the apparent admittance should tend to
\b{$\pi/4$} at high frequencies (Warburg impedance).  This value seems
in agreement with phase measurements in cerebral
cortex~\cite{gomes2016,BedDes2017a} and retina~\cite{Pham}, and is
also consistent with Kramers-Kronig relations.

A further interesting property, developed Section~\ref{sec1.3}, is
that the apparent permittivity is far from negligible at very low
frequencies when the imaginary part of the electric conductivity is
non zero, even if it is very small.  This property must be related to
the difficulty of measuring the impedance (or admittance) at low
frequencies~\cite{Gabriel,Logothetis,gomes2016,BedDes2017a}.  This
difficulty could be the sign that the apparent permittivity increases
rapidly at low frequencies (lower than $\sim$10~Hz).  This phenomenon
should not appear in a resistive medium.

We conclude that, given the constraints imposed by Kramers-Kronig
relations, the previous experimental measurements seems internally
consistent for frequencies larger than 10~Hz, either for
resistive~\cite{Logothetis,Miceli} or non-resistive
media~\cite{Gabriel,gomes2016}.  Further experiments are needed to
distinguish between these two alternatives, as reviewed in detail
previously~\cite{BedDes2017}.

\end{document}